\documentclass[]{interact}

\usepackage{epstopdf}
\usepackage[caption=false]{subfig}
\usepackage{tikz}
\usetikzlibrary{decorations.pathreplacing}
\usetikzlibrary{positioning}
\usepackage[numbers,sort&compress]{natbib}

\bibpunct[, ]{[}{]}{,}{n}{,}{,}
\makeatletter
\def\NAT@def@citea{\def\@citea{\NAT@separator}}
\makeatother

\theoremstyle{plain}

\theoremstyle{definition}

\theoremstyle{remark}

\begin{document}

\articletype{Review}

\title{Recent Advances in Maximum Entropy Biasing Techniques for Molecular Dynamics}

\author{
\name{D. Amirkulova\textsuperscript{a} and A.~D White\textsuperscript{a}\thanks{CONTACT A.~D. White. Email: andrew.white@rochester.edu}}
\affil{\textsuperscript{a} University of Rochester, Department of Chemical Engineering, Rochester, NY, USA}
}

\maketitle

\begin{abstract}
This review describes recent advances by the authors and others on the topic of incorporating experimental data into molecular simulations through maximum entropy methods. Methods which incorporate experimental data improve accuracy in molecular simulation by minimally modifying the thermodynamic ensemble. This is especially important where force fields are approximate, such as when employing coarse-grain models, or where high accuracy is required, such as when attempting to mimic a multiscale self-assembly process. The authors review here the experiment directed simulation (EDS) and experiment directed metadynamics (EDM) methods that allow matching averages and distributions in simulations, respectively. Important system-specific considerations are discussed such as using enhanced sampling simultaneously, the role of pressure, treating uncertainty, and implementations of these methods. Recent examples of EDS and EDM are reviewed including applications to {\em ab initio} molecular dynamics of water, incorporating environmental fluctuations inside of a macromolecular protein complex, improving RNA force fields, and the combination of enhanced sampling with minimal biasing to model peptides.  
\end{abstract}

\begin{keywords}
molecular dynamics; maximum entropy
\end{keywords}

\section{Introduction}

A common task in molecular simulation is ensuring that observables of the simulation match experimentally measured values. For example, in simulations of protein structure\cite{Lindorff-Larsen2010} or liquids\cite{Shivakumar2010}, quantitative agreement with experiments is the standard for assessing correctness of a model. When there is no quantitative agreement, changing the potential energy function or adding additional components to the simulation are possible ways to improve the fit. Making such changes can be an ambiguous and challenging process, especially if the potential energy function has multiple terms that can be modified. Minimal biasing techniques are a class of methods that modify a potential energy function to improve quantitative agreement with experimental values while minimizing the the change in the potential energy function. The definition of ``minimal'' and the way the potential energy function is modified vary from method to method. 

Recent reviews of minimal biasing methods can be found in \citet{Sormanni2017},  \citet{bonomi2017}, and \citet{Olsson2013}. Table 1 in \citet{bonomi2017} provides an overview of 28 minimal biasing methods, categorizing them by whether they maximize entropy, maximize parsimony, or use Bayesian inference. These three categories correspond broadly to the criteria used to ensure that the biasing function introduces a \textit{minimal} change to the potential energy function. This review focuses on two techniques developed by the authors that are categorized as maximum entropy methods: experiment directed simulation\cite{white2014efficient} (EDS) and experiment directed metadynamics\cite{White2015b} (EDM). EDS is for matching ensemble average scalars and EDM is for matching free energy surfaces (probability distributions of observables).

EDS, like other minimal biasing methods, modifies a potential energy function to change ensemble averages of observables to match a specific value. These observables are typically equivalent to collective variables, but are intended to be only those that area experimentally verifiable quantities. What separates EDS from other methods is that it does not use replicas and can be used to construct a continuous NVE trajectory. For example, the Bayesian landscape tilting method of \citet{Beauchamp2014} relies on post-processing so that there cannot be a continuous NVE trajectory. Another example is the replica method of \citet{Lindorff-Larsen2005} which relies on replica-exchange of biasing forces and thus cannot result in a continuous trajectory. This does not mean EDS is ``better''; indeed these two methods appear to provide better sampling and scaling than EDS. However, the ability to compute an NVE trajectory allows dynamic observables like hydrogen-bonding lifetimes to be computed. The key results from EDS have been to improve thermodynamic observables and indirectly improve dynamic observables. For example, EDS was recently used to create a state-of-the-art DFT water model that gives near perfect agreement with X-ray scattering results, water diffusivity, and proton-hopping behavior by improving only the water oxygen-oxygen coordination number\cite{White2017}.

EDM is a maximum entropy method that matches ensemble probability distributions, or equivalently free energy surfaces, using prescribed functions. So far, EDM has proved most useful for matching radial distribution functions (RDFs), e.g. to then do coarse-grained modeling. EDM is less often used because it is rare to have experimental data that gives a probability distribution, other than a normal distribution (which is better treated with methods like in \citet{Hummer2015}). Other groups have since arrived at the same approach as EDM and typically it is now called ``targeted metadynamics'' because it uses the method of metadynamics to arrive at a target free energy surface\cite{Marinelli2015,Gil-Ley2016}. EDM is also equivalent to variationally enhanced sampling metadynamics\cite{valsson2014variational}, although there the target is a tool to improve sampling, and the biased simulation is not the goal. The name EDM was chosen by the authors because at the time there were new metadynamics methods that could target a collective variable domain\cite{Dama2015}. Now the term ``targeted'' seems to apply exclusively to target distributions, and EDM is probably best described as falling under the umbrella of targeted metadynamics methods.

Here we review the maximum entropy derivations that underpin both EDS and EDM, describe some of the recent research benefiting from these methods, and discuss implementation details such as understanding the effect of EDS and EDM on the virial and accounting for uncertainty in the experimental data. 

\section{Theory}
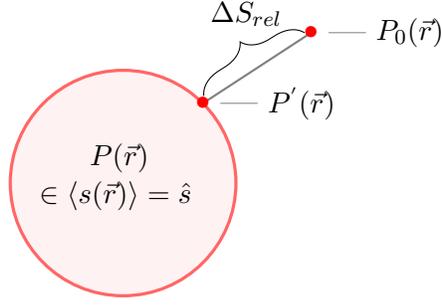
\begin{figure}
    \centering
  
\begin{tikzpicture}[scale=.5]
\filldraw[color=red!60, fill=red!5, very thick](0,0)
circle (3);
\draw[gray, thick] (2.1213203,2.1213203)--(5,4);
\node [red, pin=right:$P^{'}(\vec{r})$] at (2.1213203,2.1213203){\textbullet};
\node [red, pin=right:$P_{0}(\vec{r})$] at (5,4)  {\textbullet};
\draw [decorate,decoration={brace,amplitude=10pt},xshift=-5pt,yshift=0pt] (2.3213203,2.3213203) -- (5,4) node [black,midway,xshift=-3pt,yshift=19pt]{$\Delta S_{rel}$};
 \coordinate (A) at (-0.1,-0.2);
    \node at (A) [above = 1mm of A] {$P(\vec{r})$};
 \coordinate (A) at (-0.2,-1.2);
    \node at (A) [above = 1mm of A] {$\in \langle s(\vec{r})\rangle=\hat{s}$ };
\end{tikzpicture}
    \caption{A schematic of the minimum relative entropy derivation. $P(\vec{r})$ is the unbiased probability distribution from the unbiased potential energy. We are finding a biased probability distribution, $P\vec{r})$, that is consistent with Equation~\ref{eq:eds-constraint}.
    There is a hypersurface of possible such probability distributions. With the condition that we minimimize relative entropy, or ``distance'' in this schematic, we find a unique point in the hypersurface $P'(\vec{r})$.
    }
    \label{fig:eds_phase_space}
\end{figure}

EDS minimally modifies an ensemble so that the ensemble average of some scalar matches a desired value, such as a value obtained from an experiment. The EDS bias is minimal because the resulting biased ensemble maximizes entropy\cite{Pitera2012, PhysRev.106.620}. Maximum entropy and minimum relative entropy are equivalent approaches to derive these minimal biasing equations\cite{Roux2013c}. We will use the minimum relative entropy approach because it has an intuitive interpretation as a distance metric. Figure~\ref{fig:eds_phase_space} illustrates how the biased ensemble is one of many choices that matches our constraints, but is as close as possible to the unbiased ensemble as measured via relative entropy. 

Consider an unbiased potential energy function, $U(\vec{r})$, which has a probability distribution of $P(\vec{r})$ under the NVT ensemble, following the Boltzmann distribution: $P(\vec{r})\propto \exp\left(-\beta U(\vec{r})\right)$. $\vec{r}$ is the set of coordinate vectors of the $N$ particles of the ensemble and $\beta = \frac{1}{kT}$. $k$ is the Boltzmann constant and $T$ is absolute temperature. We would like to find a biased ensemble $P'(\vec{r})$ which is as similar as possible to the unbiased ensemble. Similarity can be defined via the relative entropy:
\begin{equation}
    \Delta S_{\textrm{rel}} = \int d\vec{r}\, P(\vec{r})\ln \frac{P(\vec{r})}{P'(\vec{r})}
\end{equation}
where $P'(\vec{r})$ is the biased ensemble probability distribution, and the integral is taken over all coordinates. Having a lower $\Delta S_{\textrm{rel}}$ means that the biased and unbiased ensembles are more similar. The biased ensemble should also have an observable average that matches a target value, which could be obtained from an experiment. This constraint is represented as
\begin{equation}
    \label{eq:eds-constraint}
   \left<s(\vec{r})\right> = \int d\vec{r}\, P'(\vec{r})s(\vec{r}) = \hat{s}
\end{equation}
where $s(\vec{r})$ is an instantaneous value for our observable (collective variable) which we are matching to $\hat{s}$, the desired scalar value. This is sometimes called a forward model and depends only on positions. We will relax this assumption below. 

These equations represent (1) a constraint (to have our average simulation value match the target $\hat{s}$), and (2) a scalar to minimize (the relative entropy). When these conditions are present, we can find the optimal value via the method of Lagrange multipliers. The Lagrangian is:

\begin{equation}
\label{eq:lagrange}
    \mathcal{L}\left[\lambda, P(\vec{r}), P'(\vec{r})\right] = \Delta S_{\textrm{rel}} - \lambda\left(\int d\vec{r}\, P'(\vec{r})s(\vec{r}) - \hat{s}\right)
\end{equation}

Equation~\ref{eq:lagrange} can be minimized by taking the functional derivative equal to zero: $\frac{\delta \mathcal{L}}{\delta P'(\vec{r})}$ = 0. Solving for $P'(\vec{r})$ gives this expression for the biased ensemble:
\begin{equation}
    P'(\vec{r}) = \frac{1}{Z'}e^{-\beta (U(\vec{r}) + \lambda s(\vec{r}))}
\end{equation}
where $Z'$ is a normalization constant. This gives an expression for the biased potential energy as $U'(\vec{r}) = U(\vec{r}) + \lambda s(\vec{r})$. This result shows that if the bias is \textit{linear} in the instantaneous observable, the ensemble is minimally biased. This derivation can be repeated for multiple dimensions\cite{Pitera2012} and for functions instead of scalars\cite{White2015b}. The general result is:
\begin{equation}
    \label{eq:bias}
    U'(\vec{r}) = U(\vec{r}) + \sum_i \lambda_i s_i(\vec{r}) + \sum_j \mu_j\left[v(\vec{r})\right]
\end{equation}
where $i$ is the index of ensemble averages that are matched to a desired value and $j$ is the index of free energy surfaces that are matched to desired functions. $\mu_j\left[v(\vec{r})\right]$ is a bias added that depends on $v(\vec{r})$, another instantaneous observable, and causes the biased ensemble to match a specific distribution in the free energy surfaces $F_t\left[v(\vec{r})\right]$. Specifically,
\begin{equation}
    \int d\vec{r}\delta(v(\vec{r}) - v')P'(\vec{r}) = q(v')
\end{equation}
where $\delta$ is the Dirac delta function and $q(v')$ is a desired probability distribution for $v(\vec{r})$ (i.e., $q[v(\vec{r})] = -\frac{1}{\beta}\ln F_t\left[v(\vec{r})\right]$). $q(v)$ could be obtained for example from a scattering or FRET experiment\cite{Boura2011}.

The derivation above gives the form of the biased potential energy that is minimally biased, but it does not enable calculation of the Lagrange multipliers. That is the purpose of the EDS and EDM methods. EDS and EDM are time-dependent methods that change the potential energy of a simulation to arrive at the bias in Equation~\ref{eq:bias}. EDS is intended to be used in a two-step process: finding the Lagrange multiplier (adaptive) and then running a standard MD simulation using the modified force field given by Equation~\ref{eq:bias}. 

During the adaptive phase of EDS, the Lagrange multipliers are called coupling constants to distinguish from the time-independent Lagrange multipliers. These are indicated as $\alpha_\tau$ where $\tau$ is a discrete step index. The Lagrange multipliers are set to be the average of $\alpha_\tau$. $\alpha_\tau$ is defined as
\begin{equation}
\label{eq:eds-update}
    \alpha_{\tau + 1} = \alpha_\tau + \eta_\tau g_\tau
\end{equation}
\begin{equation}
\label{eq:eds-gradient}
    g_\tau = 
     \frac{2\beta}{w}\left(\left<s\right>_\tau - \hat{s}\right)\left(\left<s^2\right>_\tau - \left<s\right>^2_\tau\right)
\end{equation}
where $w$ is an arbitrary constant used to ensure unit homogeneity, $\left<\cdot\right>_\tau$ is the ensemble average between step $\tau - 1 $  and $\tau$, and $\eta_t$ is 
\begin{equation}
    \eta_\tau = \frac{A}{\sqrt{\sum_i^{\tau} g_i^2}}
\end{equation}
where $A$ is a user-defined constant that controls the size of the first step. The point of $\eta_\tau$ is to reduce the size of the steps over time. Note that because $g_{\tau}$ is in the sum, $\left|\alpha_0\right| = A$. Typically the gradients should be clipped for stability and $A$ is a natural upper/lower bound for clipping it. Thus a user of this method must choose $A$ and the time between updates. This update procedure is derived in \citet{white2014efficient} and is based on per-coordinate infinite horizon stochastic gradient descent\cite{McMahan2010}. \citet{Hocky2017} recently found that Equation~\ref{eq:eds-gradient} can be modified to use covariance in multiple dimensions to improve convergenece and that replacing Equation~\ref{eq:eds-update} with Levenberg--Marquardt optimization further improves convergence.

The EDM method finds the $\mu\left[v(\vec{r})\right]$ bias function in Equation~\ref{eq:bias}. For compactness, we will now omit the dependence on $\vec{r}$. Unlike EDS, EDM consists of a single phase where $\mu(v)$ changes less over time. The update equation is
\begin{equation}
    \label{eq:edm-update}
    \mu(v)_{\tau+1} = \mu(v)_\tau + \frac{1}{q(v_\tau)}\exp({-\theta(v_\tau, v)_\tau})G(v, v_\tau)
\end{equation}
where $v_\tau$ is the value of $v(\vec{r})$ at time $\tau$, $G$ is a kernel function (e.g., Gaussian), $q(v_\tau)$ is the probability at position $v_\tau$ from the target probability distribution and $\theta(v_\tau, v)_\tau$ is a function which controls convergence. Following the above update rule causes the simulation to converge to the following distributions\cite{White2015b, Dama2014a}, depending on the choice of $\theta$:

\begin{equation}
    \label{eq:edm-converge}
    \theta(v_\tau, v)_\tau =\left\{\begin{array}{lr}
    1, & \textrm{non-convergent} \\
    \hat{\mu}_\tau, & q(v)\\
    \beta \Delta T \mu(v_\tau)_\tau / T, & q(v)^{\Delta T / (\Delta T + T)} P(v)^{T / (\Delta T + T)}
    \end{array}\right.
\end{equation}
where $\Delta T$ is the ``tempering factor''\cite{White2015b}, $P(v)$ is the marginal unbiased distribution for $v(\vec{r})$, and $\hat{\mu}$ is the total or average of all previous $\mu_\tau$. The first condition, $\theta = 1$, is similar to normal metadynamics; it reaches a distribution similar to $q(v)$ but may oscillate due to a lack of dampening in the update size. Condition two, called globally tempered, converges correctly to $q(v)$. The last condition, locally tempered, converges to an adjustable mixture of the unbiased and target distribution. EDM is traditionally implemented as globally tempered, so that the the final distribution is indeed the target. The first condition is good for tuning parameters, since it makes progress more quickly. The final condition, locally tempered, allows a pseudo-Bayesian tuning of prior belief in the unbiased ensemble. It is pseudo-Bayesian because the ratio of influence from the prior belief to evidence is computed, not set, in Bayesian modeling.

An important consideration of both EDS and EDM is that they add potential energy during the update step which quickly becomes kinetic energy. Thus, it is important that the thermostat used to maintain constant temperature in the NVT can dissipate energy faster than it is added by the update steps. This is necessary during the adaptive phase of EDS and EDM prior to convergence.

\subsection{Treating Uncertainty in Experimental Data}
One complication of minimal biasing methods is uncertainty in the experimental data. For example, the experimental data could be the radius of gyration of a polymer with a reported uncertainty in the mean of 5 nm. Should the radius of gyration be matched exactly, or only to within 5 nm? It is possible to stop the adaptive phase of EDS early, for example when the average is within the uncertainty of the experimental data. An early stop is ad-hoc and part of the maximum entropy derivation unlike other methods which are built to address uncertainty\cite{Hummer2015,Brookes2016}.

\citet{Cesari2016} proposed a modification to the maximum entropy derivation above to address experimental uncertainty. Instead of the constraint in Equation~\ref{eq:eds-constraint}, this constraint is used
\begin{equation}
    \label{eq:bussi-constraint}
    \int d\vec{r}\, P''(\vec{r}, \epsilon)\left[s(\vec{r}) + \epsilon\right] = \hat{s}
\end{equation}
where $\epsilon$ is an auxiliary variable that allows deviations in the average of $s(\vec{r})$ and $P''(\cdot)$ is the probability distribution that will be solved after maximizing entropy. $\epsilon$ is a random variable from a prior distribution $P_0(\epsilon)$ that describes the uncertainty in the experimental data. For example, $P_0(\epsilon)$ could be a normal distribution or a Laplace distribution. \citet{Cesari2016} show that $P''(\vec{r}, \epsilon) = 1 / Z'' P'(\vec{r})P_0(\epsilon)e^{\lambda\epsilon}$. This leads to a different update step of 
\begin{equation}
    \label{eq:bussi-update}
        g_\tau = 
     \frac{2\beta}{w}\left(\left<s\right>_\tau + \xi(\alpha_\tau) - \hat{s}\right)\left(\left<s^2\right>_\tau + \frac{\partial \xi(\alpha_\tau)}{\partial\alpha} - \left<s\right>^2_\tau\right)
\end{equation}
where $\xi(\alpha_\tau)$ is the analytic posterior of $\epsilon$:
\begin{equation}
    \xi(\alpha_\tau) = \frac{\int d\epsilon\, P_0(\epsilon) e^{-\alpha_\tau \epsilon}\epsilon}{\int d\epsilon\, P_0(\epsilon) e^{-\alpha_\tau \epsilon}}
\end{equation}
Returning to the radius of gyration example above, if the uncertainty ($P_0(\epsilon)$ is assumed to be a normal distribution normal with $\sigma = 5$nm, then $\xi(\alpha_\tau)= -25\alpha_\tau$. This new update-step can be used to rigorously include experimental uncertainty in to the EDS method.

EDM is able to tune the relative importance of the target distribution and the unbiased ensemble with the locally-tempered variant in Equation~\ref{eq:edm-converge}. This could be used to match intuition. For example, if you believe that the target distribution which comes from experimental data is twice as accurate as the molecular dynamics simulation, you could choose $\Delta T / (\Delta T + T) = 0.66$ giving about twice as much weight to the target distribution. 

\section{Applications of EDS and EDM}

\begin{figure}
    \centering
    \includegraphics[width=5cm]{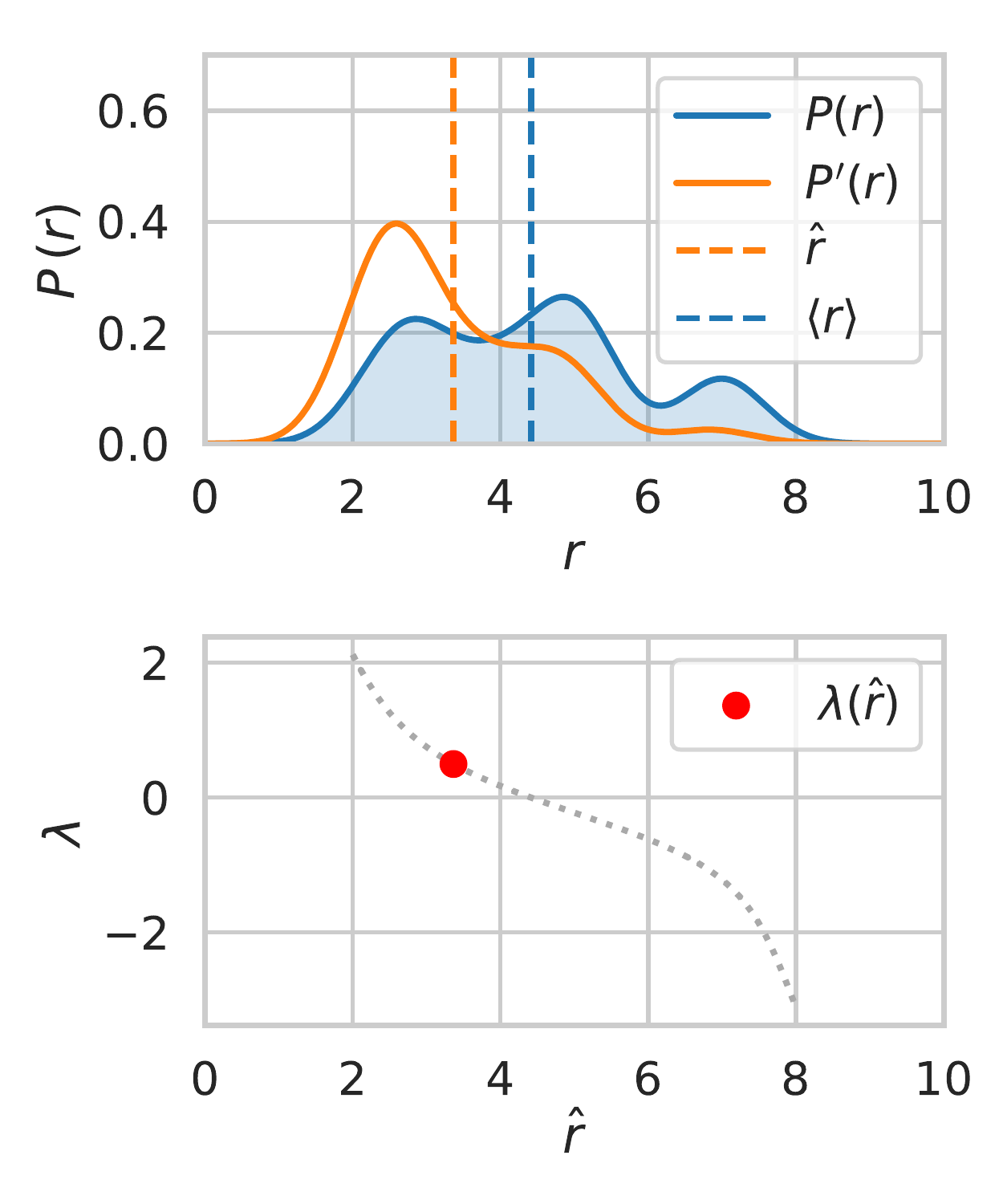}
    \caption{A 1-D EDS calculation where the mean of a probability distribution function (PDF) is being biased to match the dash vertical orange line. The top plot shows the unbiased and biased PDFs. The biased PDF shows as much of the shape of the unbiased PDF as possible while matching the new biased mean. The bottom shows the Lagrange multipliers that give all possible biased means. The red dot indicates the current Lagrange multiplier for the biased PDF. There is a unique $\lambda$ for all possible biased means, as discussed in the Theory section. }
    \label{fig:eds-model}
\end{figure}

A model 1-D system is shown in Figure~\ref{fig:eds-model} as a probability distribution function, $P(r)$. EDS is being used to modify the average value of $r$ to match a new set point, $\hat{r}$. EDS adds a linear bias, whose strength is indicated with the red dot, to create the biased PDF $P'(r)$ according to Equation~\ref{eq:bias}. Notice how the features of $P(r)$ are mostly maintained in $P'(r)$. The bottom plot shows how each value of $\hat{r}$ corresponds to a unique biasing strength.

\begin{figure}
    \centering
    \includegraphics[width=5cm]{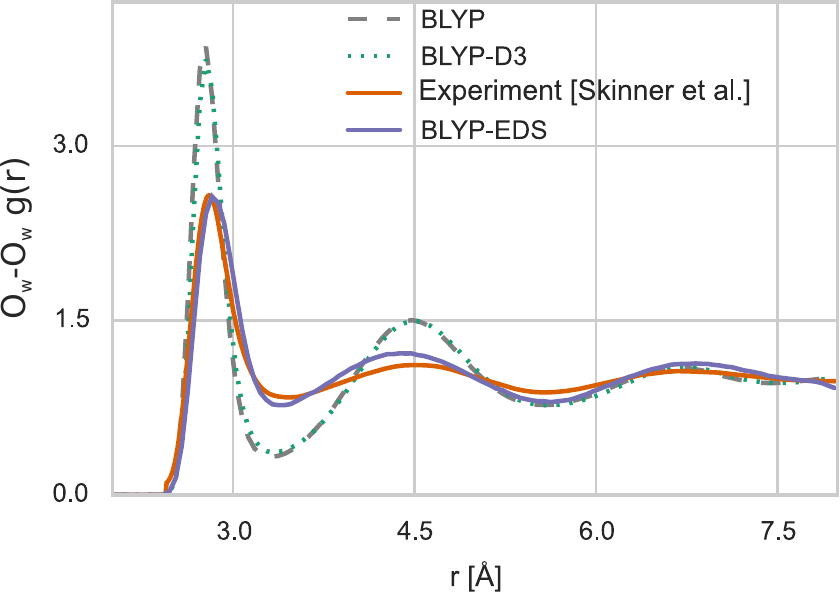}
    \caption{Water oxygen-oxygen radial distribution function from {\em ab inito} molecular dynamics at 300 K NVT and experiments from \citet{Skinner2013}. BLYP and BLYP-D3 are from DFT with and without dispersion corrections with the BLYP exchange functional. Note that BLYP-D3 is typically done at 330K, which gives much better agreement\cite{Tse2015}. See \citet{White2017} for complete system details. The BYLP-EDS line is DFT with EDS bias added to the water oxygen-oxygen coordination number. EDS shows near quantitative agreement with experiment. Copyright 2017 AIP Publishing LLC. }
    \label{fig:eds-aimd}
\end{figure}

A more sophisticated system which demonstrates the capabilities of improving dynamic observables is the recent work on EDS {\em ab initio} molecular dynamics (AIMD) simulations of water\cite{White2017}. DFT water with the BLYP exchange functional poorly represents water structure as seen in Figure~\ref{fig:eds-aimd} (black line). It is over-structured and has water self-diffusion coefficients that are too high (0.005-0.005 \AA$^2$ / ps)\cite{Tse2015} compared with the experimental value of 0.23 \AA$^2$\cite{DC9786600199}. EDS was used to improve the coordination number of the oxygen-oxygen (O$_w$-O$_w$) water molecules and resulted in near perfect agreement with experimental scattering data (Figure~\ref{fig:eds-aimd}). A number of unrelated observables improved as well, including RDFs and the water self-diffusion coefficient, which increased to 0.06$\pm\,$\AA$^2$/ps. \citet{White2017} further demonstrated that the EDS bias could be transferred to excess proton-water simulations and that when combined with DFT dispersion corrections\cite{Sitarz2000}, the agreement further improves. The accuracy improvement with the dispersion corrections shows that EDS is general in its applicability to DFT methods.

\citet{Cortina2018} used EDS to study the KPC-2 carbapenemase enzyme, which is responsible for drug resistance in the majority of carbapenem-resistant Gram-negative bacteria\cite{doi:10.1021/acs.jmedchem.7b00158}. They used EDS to modify protein-protein distances while doing a committor analysis\cite{du1998transition} to identify transition states in the carbapenem-enzyme complex. \citet{Cesari2016} used a method similar to EDS (modified update step) to improve agreement with experimental NMR $^3$J coupling data for RNA oligonucleotids. After biasing, they used the simulation results to improve the underlying Amber force field and thus create a transferrable model. \citet{Cesari2016} also developed a novel approach to account for experimental data uncertainty by adding auxiliary variables to the EDS update step. 

EDM has been used less than EDS due to the rarity of experimental data giving an exact probability distribution. One example explored in the original EDM paper was to construct a mean field bias that mimics an alanine dipeptide being in the backbone of a protein\cite{White2015b}. This was done by computing a potential of mean force (PMF) for $\phi,\psi$ dihedral angles from PDB crystallography data for the alanine-alanine sequence in protein structures. This corresponds to a probability distribution and a molecular dynamics simulation of alanine dipeptide was done with its $\phi,\psi$ dihedral angles biased to match the desired values. The result was a molecular dynamics simulation where the dipeptide was biased to behave as part of a longer protein structure. A similar approach was later used by \citet{Hocky2017} with EDS to built a pseudo-mean field model for actin filaments. 

Another example of EDM can be found in \citet{Gil-Ley2016} who used it (called targeted metadynamics) to improve agreement of RNA oligonucleatides' dihedral angles with PDB crystallography data. They similarly built dihedral angle PMFs using the crystallography data and  biased molecular dynamics simulations of RNA to adopt the dihedral angle probability distribution functions. \citet{Gil-Ley2016} then took the converged EDM bias and transferred it to a different simulation of an RNA tetramer and found little to no improvement of agreement with the crystallography data. This may have been due to the underlying assumption that a PMF derived from crystallographic data is not representative of the room temperature ensemble. Nevertheless, EDM is a promising technique for incorporating experimental data as a type of auxiliary potential energy function in simulations.

\subsection{Coarse-Grain Modeling}
EDS and EDM are well-suited for coarse-grain (CG) modeling because they guarantee that a CG model matches experimental data. \citet{Dannenhoffer-Lafage2016} showed that EDS can be applied before force-matching (a CG method) and the improvements of agreement with experimental data is maintained. \citet{Dannenhoffer-Lafage2016} began with all atom molecular dynamics simulations of Ethylene carbonate with the force field from \citet{Masia2004}, which is known to not match center-of-mass coordination numbers from more accurate DFT calculations from \citet{Borodin2009}. \citet{Dannenhoffer-Lafage2016} biased the all-atom simulations with EDS to match these known coordination numbers. They then used force-matching to create multiple CG models with one, two, and three site CG beads, using either biased all-atom or unbiased all-atom simulation data. The biased all-atom simulations showed better agreement with coordination numbers, indicating that the improvement in all-atom simulations translates to improvement in the CG model with EDS. Of course, EDS could be used directly on the CG model but that can lead to complications in how observables are calculated on CG models\cite{Wagner2016b}.

EDS has been applied to CG simulations of the G- and F-actin proteins as monomers and trimers\cite{Hocky2017}. Actin proteins are globular proteins that can polymerize into long semi-flexible filaments. Their hypothesis was that they could model a subsystem from within the polymerized actin structure by incorporating information about structural fluctuations from simulations of the larger system via EDS. By biasing the first and second moment of two important collective variables in actin monomer structure, they were able to observe filament-like conformations, and importantly, the fluctuations of these observables for an actin monomer, even in the system which contained only a single monomer solvated in water. \citet{Hocky2017} also studied a number of questions about the EDS method in their system and found the following conclusions: (1) the linear term of EDS better matches target values and maintains system fluctuations than harmonic biases; (2) replacing the variance term in Equation~\ref{eq:eds-gradient} with a covariance matrix for all biased dimensions has faster convergence; (3) Levenberg--Marquardt\cite{Stinis2005} converges faster than stochastic gradient descent. This paper also derived a simple equation that can be used to guess the value of $\lambda$ without doing a stochastic minimization (whose accuracy depends on the distance of the target observables from the unbiased observables), and hence can serve as a good initial guess for starting an EDS simulation.

\subsection{Enhanced Sampling}
EDM may require enhanced sampling if there are slow degrees of freedom orthogonal to the biased collective variable. This is most conveniently treated via the extensive literature on enhanced sampling with metadynamics, since EDM is a type of metadynamics and typically implemented within a metadynamics code. EDS is not as simple because it requires that the $\left<\cdot\right>_\tau$ term in Equation~\ref{eq:eds-gradient} be taken over an NVT ensemble. This limits the enhanced sampling techniques to those that still give correct NVT ensemble averages. One example is parallel-tempering replica-exchange. It is possible to use metadynamics if an appropriate estimator\cite{Tiwary2015} is done to compute the averages but this has not been explored in practice. 

\citet{Amirkulova2018} demonstrated the use of enhanced sampling and EDS with the parallel-tempering well-tempered ensemble (PT-WTE)\cite{Bonomi2010}. The PT-WTE method is a enhancement of enhanced parallel-tempering replica-exchange that improves exchange rates and reduces the required number of replicas\cite{Deighan2012a}. PT-WTE satisfies the observable that the ensemble averages, $\left<\cdot\right>_\tau$, can be computed during the course of the simulation because PT-WTE only changes the magnitude of potential energy fluctuations, not their expectations. One apparent drawback is that the method loses the one-replica observable of EDS that allows computation of dynamic observables. However, this only applies to the adaptive phase. During the fixed-bias second phase of EDS, one replica can again be used to allow analysis of dynamic observables.  

\begin{figure}
    \centering
    \includegraphics[scale=0.45]{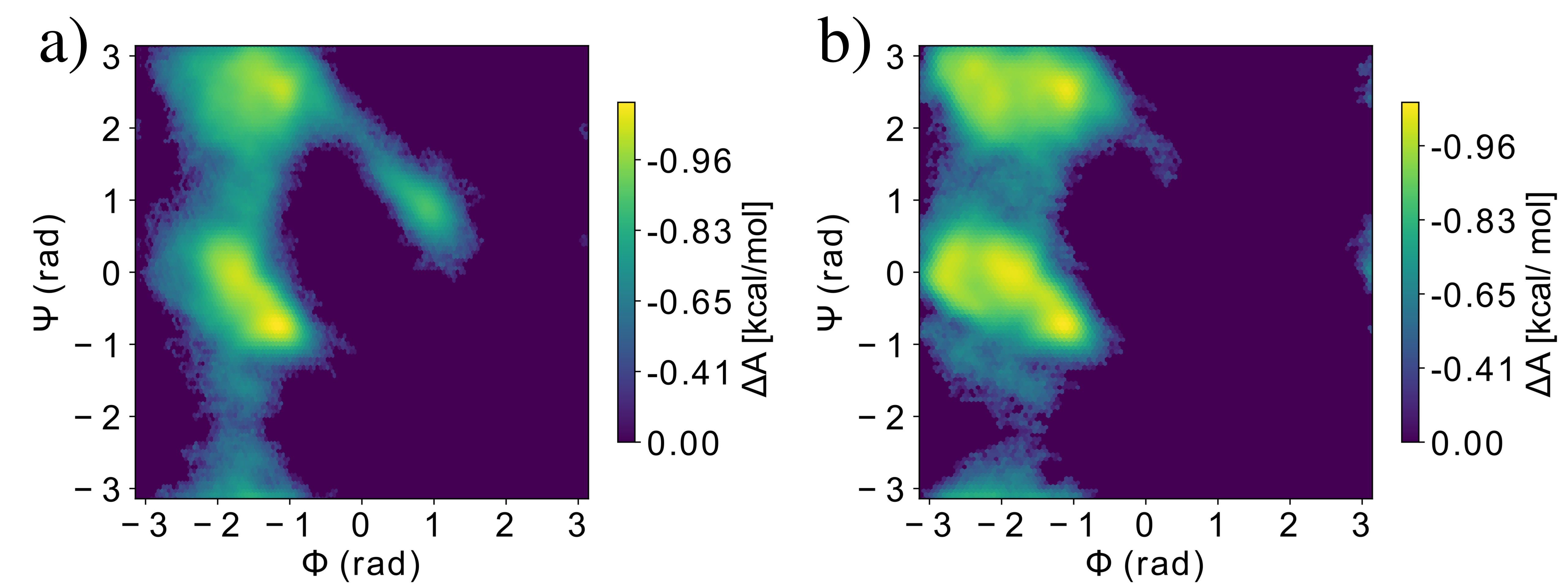}
    \caption{Free energy surface of GYG peptide along dihedral angles (Ramachandran plot) from \citet{Amirkulova2018}. a) EDS and enhanced sampling with PT-WTE and b) no EDS and no enhanced sampling. The difference between panels a and b shows that EDS changes the global free energy minimum, which better matches data from \citet{Ting2010}, and PT-WTE improves sampling based on explored regions. Copyright 2018 World Scientific Publishing Company.}
    \label{fig:gyg_fes}
\end{figure}

\citet{Amirkulova2018} studied the GYG peptide with the EDS plus enhanced sampling approach. Eight simulations were conducted, each with 1, 8, or 16 replicas.  The simulations had EDS, PT-WTE, and/or parallel-tempering. EDS was used to bias proton chemical shifts to improve agreement with experimental NMR data. The simulation results showed that PT-WTE improves sampling in the EDS method and does not change agreement with experimental data. PT-WTE also converged the EDS bias with fewer replicas than the PT method. One consideration in all EDS simulations is the effect on unbiased observables. The Ramachandran plots of the simulations with and without EDS bias are compared in Figure ~\ref{fig:gyg_fes}. When EDS and enhanced sampling are used (Figure~\ref{fig:gyg_fes} a)), the simulation explores a larger region of configurational space relative to the control simulation (Figure~\ref{fig:gyg_fes} b)). Also, the global minimum changes when using EDS in Figure~\ref{fig:gyg_fes}, bringing it closer to what was found in \citet{Ting2010} for the GYG sequence.

\section{The Role of Pressure}
Thus far we have discussed NVT and NVE ensembles. EDM and EDS add new forces to the simulation and thus affect the system virial. This can lead to undesirable density changes when the bias is subsequently used in an NPT simulation. It is possible to rectify this change in virial, and thus density, by adding a further constraint that the virial be unchanged while still maximizing entropy. As shown in the appendix, this leads to an unsolvable equation and thus it is not possible to simultaneously set the virial and other constraints. One intuitive reason for this is that the virial theorem requires the average virial be proportional to average temperature. Therefore if we tried to bias an ensemble so that its virial is fixed, then the average temperature would change, violating our constant average temperature.

One way around this challenge lies in the correlation between biased observables in EDS. When biasing multiple observables which are correlated, there is in fact no longer a unique set of Lagrange multipliers\cite{Pitera2012}. These extra degrees of freedom in the choice of Lagrange multipliers that maximize Equation~\ref{eq:lagrange} enable us to choose the Lagrange multipliers that minimally change the virial. Equation~\ref{eq:eds-gradient} can be modified so that our coupling constants minimize the additional pressure which they exert\cite{Louwerse2006}:
\begin{equation}
    \label{eq:eds-virial-gradient}
    g_\tau = 
     \frac{2\beta}{w}\left(\left<s\right>_\tau - \hat{s}\right)\left(\left<s^2\right>_\tau - \left<s\right>^2_\tau\right) + 2\nu \Delta p_\tau\frac{\partial \Delta p_\tau}{\partial  \alpha_\tau}
\end{equation}
\begin{equation}
    \Delta p_\tau = -\frac{\alpha_\tau}{w}\frac{ds}{dV} = -\frac{\alpha_\tau}{w}\sum_{ij}\left(\frac{\partial s}{\partial v_{ij}}\frac{\partial v_{ij}}{\partial V}\right)
\end{equation}
where $\nu$ is a parameter that controls the importance of minimizing the virial, $\Delta p_\tau$ is the change in the system virial pressure due to the EDS bias, $\frac{\partial s}{\partial v_{ij}}$ is the partial derivative of the observable with respect to one component of the triclinic box matrix, and $\frac{\partial v_{ij}}{\partial V}$ can be computed from the adjugate of the triclinic box matrix. When biasing multiple observables, $\Delta p_\tau$ should be the total over all observables. $g_\tau$ is in units of $s^2$ per energy, so $\nu$ must be in units of volume squared times $s^2$ per energy squared. Practically we can choose a unitless $\nu^*$, which is defined from
\begin{equation}
\nu = \nu^* \frac{w^2\beta^2}{\rho^2}
\end{equation}

\begin{figure}
    \centering
    \includegraphics[width=\textwidth]{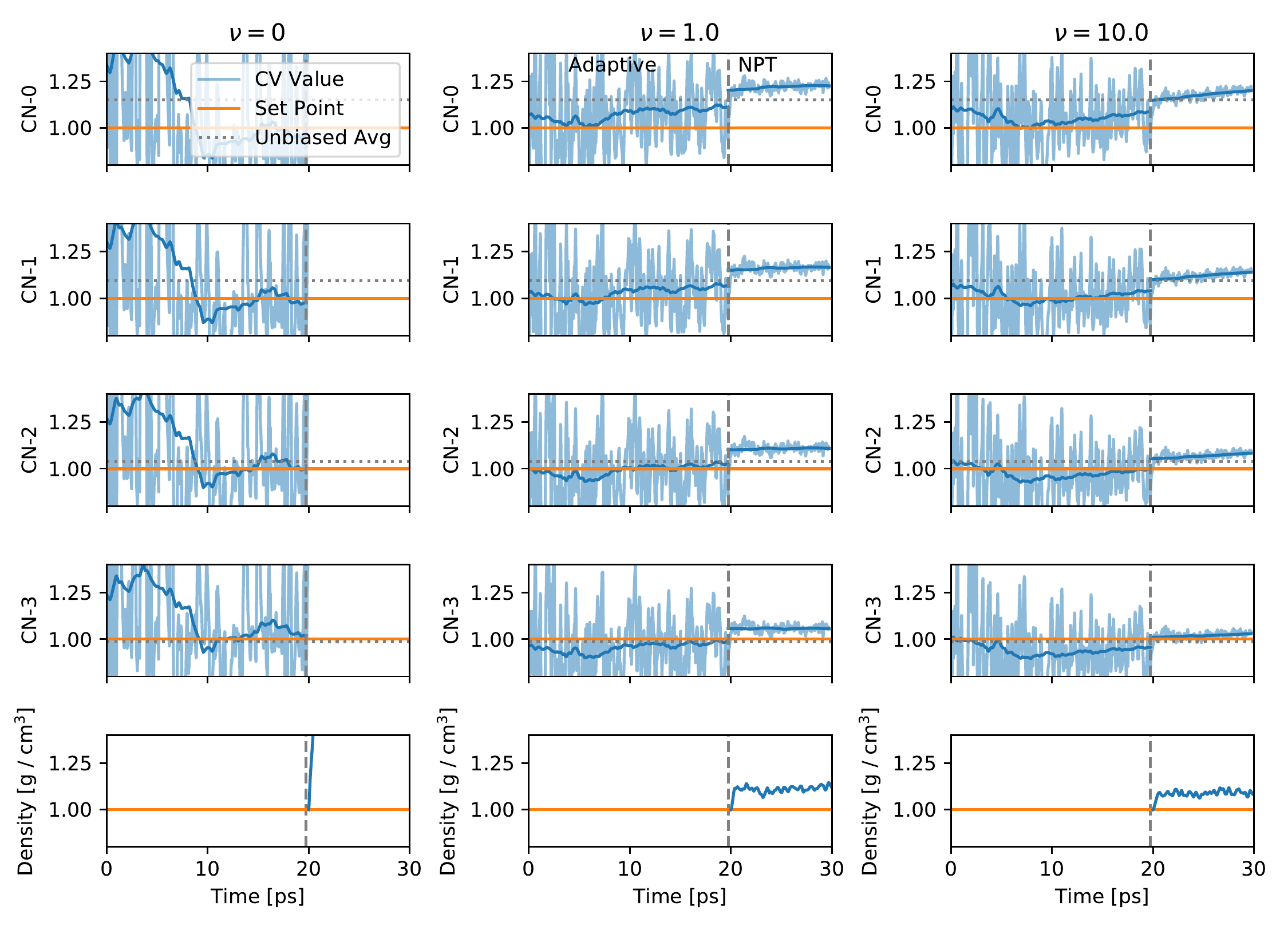}
    \caption{EDS applied to water coordination number moments 0 through 3 in molecular dynamics simulation of over-structured SCP/E water model. The columns show increasing strength of virial minimization. Each plot is collective variable scaled by its set-point. The vertical dashed line spearates the NVT adaptive simulation and a fixed bias NPT phase. The results show that increasing the strength of the virial minimization actually improves convergence by reducing large-magnitude changes in biasing force. Without virial minimization, EDS can produce nonphysical densities when the NVT bias is transferred to NPT. The change in density moves the CVs. For $\nu^* = 0$, this meant CV values beyond the y-limits of the plot.}
    \label{fig:virial-cvs}
\end{figure}

Equation~\ref{eq:eds-virial-gradient} was used in a molecular dynamics simulation of 128 modified SPC/E water molecules. This modified SPC/E water has increased charges ($q_O = -0.94$) to distort its coordination number. EDS was then applied to correct the coordination number using experimentally derived coordination numbers from \citet{Skinner2013} with varying strengths of the virial correction. The coordination number moment definition may be found in \citet{white2014efficient}. Figure~\ref{fig:virial-cvs} shows that coordination number and its moments are still correctly biased with the virial term. The EDS parameters were a range ($A$) of 50 kj/mol, a period of 25 fs, and the Levenberg--Marquardt optimization procedure. A Nose-Hoover thermostat with a time constant of 25 fs and a timestep of 0.5 fs were used for the molecular dynamics in the LAMMPS simulation engine\cite{Plimpton1995}. The NPT barostat was Parrinello-Rahman.

\begin{figure}
    \centering
    \includegraphics[width=\textwidth]{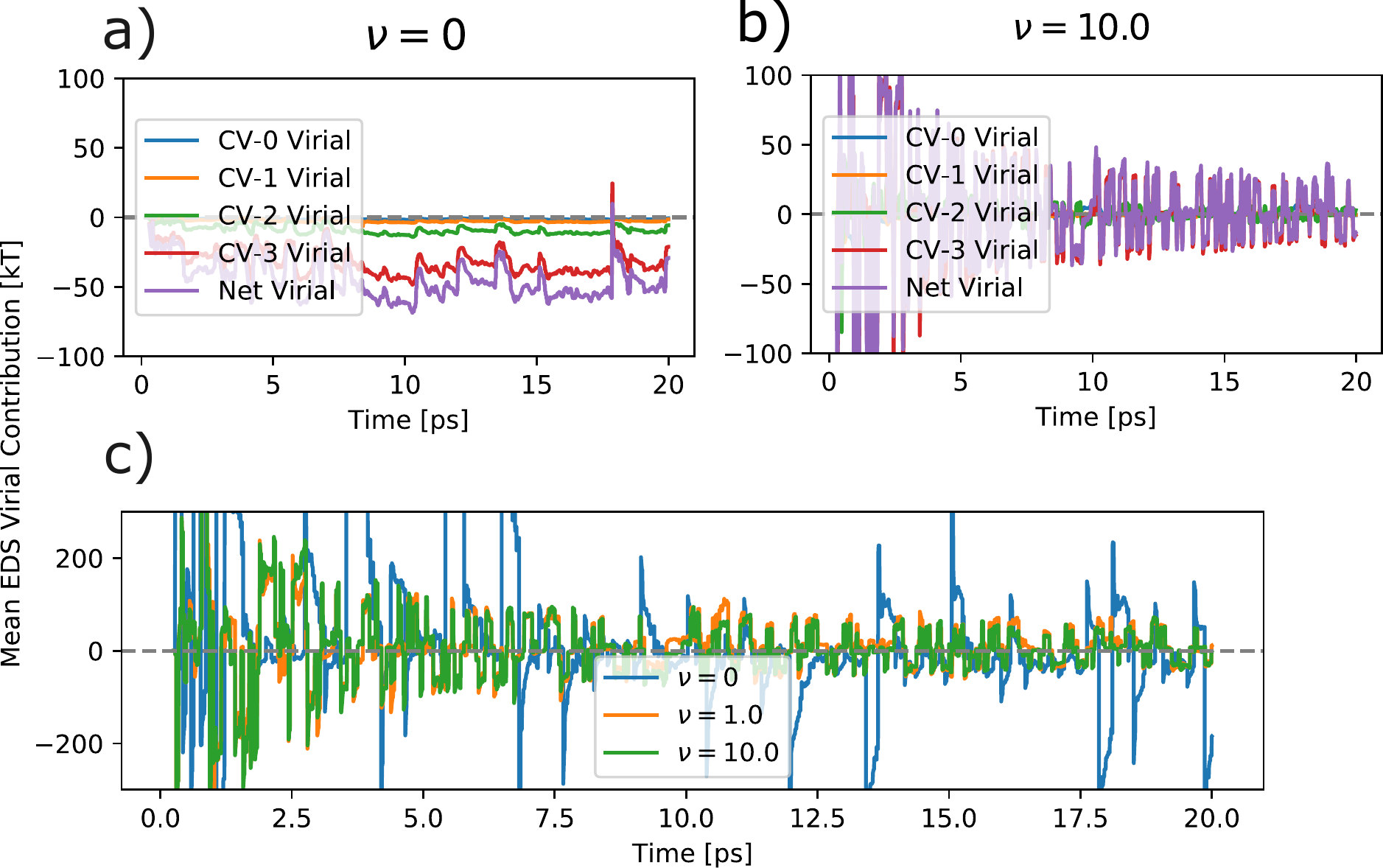}
    \caption{The EDS contribution to virial as EDS is applied to water coordination number moments 0 through 3 in molecular dynamics simulation of over-structured SCP/E water model. Panels a and b compare the virial contribution in a normal EDS and EDS with virial minimization strength of $\nu^* = 1$. There is lower per-collective variable virial contribution with virial minimization. Panel c shows the effect of different $\nu^*$ values on the total virial contribution from EDS.}
    \label{fig:virial-virial}
\end{figure}

Figure~\ref{fig:virial-virial} shows the effect of the virial penalty term on the EDS virial contribution. The virial contribution plotted and computed here is the mean per-particle virial energy contribution. That is $\Delta p_\tau / \rho / kT$ which removes the effect of particle number and energy scale. Panel a shows the virial contribution of each collective variable with $\nu^* = 0$ (no virial minimization). The net average virial contribution is 39.2 kT. Panel b shows $\nu^* = 10$ and there is a much lower contribution of 5.02 kT. Panel c compares the net virial contribution of three different $\nu^*$ values.

Figure~\ref{fig:virial-cvs} shows the impact on the convergence of the biased coordination number and its moments with the virial minimizing terms. Interestingly, adding a virial minimization term actually improves convergence. This is due to the well-known effect of regularization, which improves convergence in optimization\cite{scholkopf2001learning}. The virial minimization term is proportional to $\alpha_\tau$, so it brings down the magnitude of $\alpha_\tau$. This prevents the large swings seen in the $\nu=0$ system. Thus, adding the virial minimization term not only reduces the virial contribution but can improve convergence by minimizing magnitudes of the coupling constants. The true test can be seen in the NPT portion past 20ps where the bias is fixed but the box dimensions are free. EDS with $\nu^* = 0$ gives poor densities in NPT with a bias computed in NVT. The virial minimization reduces the change in density with a trade-off in match to the bias set-points. Virial minimization is in general recommended due to its enhanced convergence and better representation of virial forces.

\section{Implementations}

There are three implementations of EDS available: Colvars\cite{COLVARS}, Plumed1.3\cite{PLUMED}, and Plumed2\cite{TRIBELLO2014604}. The Colvars and Plumed1.3 implementations match the original EDS manuscript. The Plumed2 implementation is actively maintained as a plugin\cite{bussi2018analyzing} and has features from more recent EDS articles\cite{Hocky2017}. For example, the covariance term in Equation~\ref{eq:eds-gradient} can be computed using the full sample covariance matrix or the sample variance. The update steps can also be computed using the Levenberg--Marquardt method\cite{Stinis2005, Hocky2017}, instead of Equation~\ref{eq:eds-update}. The Plumed2 implementation is recommended. 

EDM is implemented in Plumed2 as well, under the ``targeted metadynamics'' keyword within the metadynamics biasing class. One of the common use-cases for EDM is to bias RDFs, which is not exactly a collective variable. RDFs are ``function collective variables'' in the sense that there is a distribution at each timestep. Thus the bias update equation, shown in Equation~\ref{eq:edm-update}, applies to each observed point in the RDF at each step (see \citet{White2015b} for equations). This can lead to thousands of updates per timestep on even small systems. To improve performance and scaling, we have created an implementation in LAMMPS that is more tightly integrated with the simulation engine than Plumed2 as described in \citet{White2015b}. Note that the addition of hills is parallelized here, which is unusual and enables the scaling as a function of particle size. Hill addition is 90\% of the CPU utilization. This was the implementation used in \citet{White2015b} for an ethylene carbonate electrolyte simulation and Lennard-Jones benchmark systems.

\section{Conclusions}
Minimal biasing techniques are an emerging area that improve the accuracy of molecular simulation and better utilize comparable experimental data. EDS and EDM are maximum entropy techniques that minimally bias an ensemble so that average observables or probability distributions of observables match set values. EDS and EDM both converge the potential to Equation~\ref{eq:bias}, allowing an NVE simulation with fixed potential energy in a single replica so that dynamic observables can be computed. EDS and EDM can bias multiple observables simultaneously and be combined with enhanced sampling methods. It is possible to treat uncertainty in experimental data using methods from \citet{Cesari2016}. Explicitly setting pressure is not possible in maximum entropy biasing, but in EDS it is possible to minimize the change to pressure when biasing multiple collective variables. This minimization can actually improve convergence of EDS by acting as a regularization. A variety of example systems have been presented here and there are implementations available in most simulation engines for both EDS and EDM.

\section*{Acknowledgement(s)}

The authors thank Prof. Pengfei Huo for reviewing the appendix mathematics, Rainier Barrett for help preparing the manuscript and Prof. Glen Hocky for assistance in surveying the literature and preparing the manuscript.

\section*{Funding}

This work was supported by the National Science Foundation CBET Div Of Chem, Bioeng, Env, \& Transp Sys (Grant \#1751471).

\section{References}

\bibliographystyle{unsrtnat}
\bibliography{minbias-review}

\begin{thebibliography}{48}
\providecommand{\natexlab}[1]{#1}
\providecommand{\url}[1]{\texttt{#1}}
\expandafter\ifx\csname urlstyle\endcsname\relax
  \providecommand{\doi}[1]{doi: #1}\else
  \providecommand{\doi}{doi: \begingroup \urlstyle{rm}\Url}\fi

\bibitem[Lindorff-Larsen et~al.(2010)Lindorff-Larsen, Piana, Palmo, Maragakis,
  Klepeis, Dror, and Shaw]{Lindorff-Larsen2010}
Kresten Lindorff-Larsen, Stefano Piana, Kim Palmo, Paul Maragakis, John~L.
  Klepeis, Ron~O. Dror, and David~E. Shaw.
\newblock {Improved side-chain torsion potentials for the Amber ff99SB protein
  force field}.
\newblock \emph{Proteins.}, 78\penalty0 (8):\penalty0 1950--1958, 2010.
\newblock ISSN 08873585.
\newblock \doi{10.1002/prot.22711}.

\bibitem[Shivakumar et~al.(2010)Shivakumar, Williams, Wu, Damm, Shelley, and
  Sherman]{Shivakumar2010}
Devleena Shivakumar, Joshua Williams, Yujie Wu, Wolfgang Damm, John Shelley,
  and Woody Sherman.
\newblock {Prediction of absolute solvation free energies using molecular
  dynamics free energy perturbation and the opls force field}.
\newblock \emph{J. Chem. Theory Comput.}, 6\penalty0 (5):\penalty0 1509--1519,
  2010.
\newblock ISSN 15499626.
\newblock \doi{10.1021/ct900587b}.

\bibitem[Sormanni et~al.(2017)Sormanni, Piovesan, Heller, Bonomi, Kukic,
  Camilloni, Fuxreiter, Dosztanyi, Pappu, Babu, Longhi, Tompa, Dunker, Uversky,
  Tosatto, and Vendruscolo]{Sormanni2017}
Pietro Sormanni, Damiano Piovesan, Gabriella~T Heller, Massimiliano Bonomi,
  Predrag Kukic, Carlo Camilloni, Monika Fuxreiter, Zsuzsanna Dosztanyi,
  Rohit~V Pappu, M~Madan Babu, Sonia Longhi, Peter Tompa, A~Keith Dunker,
  Vladimir~N Uversky, Silvio C~E Tosatto, and Michele Vendruscolo.
\newblock {Simultaneous quantification of protein order and disorder}.
\newblock \emph{Nat. Chem. Biol.}, 13\penalty0 (4):\penalty0 339--342, 2017.
\newblock ISSN 1552-4450.
\newblock \doi{10.1038/nchembio.2331}.

\bibitem[Bonomi et~al.(2017)Bonomi, Heller, Camilloni, and
  Vendruscolo]{bonomi2017}
Massimiliano Bonomi, Gabriella~T. Heller, Carlo Camilloni, and Michele
  Vendruscolo.
\newblock {Principles of protein structural ensemble determination}.
\newblock \emph{Curr. Opin. Struct. Biol.}, 42:\penalty0 106--116, 2017.
\newblock ISSN 1879033X.
\newblock \doi{10.1016/j.sbi.2016.12.004}.

\bibitem[Olsson et~al.(2013)Olsson, Frellsen, Boomsma, Mardia, and
  Hamelryck]{Olsson2013}
Simon Olsson, Jes Frellsen, Wouter Boomsma, Kanti~V Mardia, and Thomas.
  Hamelryck.
\newblock {Inference of structure ensembles of flexible biomolecules from
  sparse, averaged data.}
\newblock \emph{PLoS One}, 8\penalty0 (11):\penalty0 e79439, 2013.
\newblock ISSN 1932-6203.
\newblock \doi{10.1371/journal.pone.0079439}.

\bibitem[White and Voth(2014)]{white2014efficient}
Andrew~D White and Gregory~A Voth.
\newblock {Efficient and minimal method to bias molecular simulations with
  experimental data}.
\newblock \emph{J. Chem. Theory Comput.}, 10\penalty0 (8):\penalty0 3023--3030,
  2014.

\bibitem[White et~al.(2015)White, Dama, and Voth]{White2015b}
Andrew White, James Dama, and Gregory~a. Voth.
\newblock {Designing Free Energy Surfaces that Match Experimental Data with
  Metadynamics}.
\newblock \emph{J. Chem. Theory Comput.}, pages 2451--2460, 2015.
\newblock ISSN 1549-9618.
\newblock \doi{10.1021/acs.jctc.5b00178}.

\bibitem[Beauchamp et~al.(2014)Beauchamp, Pande, and Das]{Beauchamp2014}
Kyle~A Beauchamp, Vijay~S Pande, and Rhiju. Das.
\newblock {Bayesian Energy Landscape Tilting: Towards Concordant Models of
  Molecular Ensembles.}
\newblock \emph{Biophys. J.}, 106\penalty0 (6):\penalty0 1381--1390, 2014.
\newblock ISSN 0006-3495.
\newblock \doi{10.1016/j.bpj.2014.02.009}.

\bibitem[Lindorff-Larsen et~al.(2005)Lindorff-Larsen, Best, DePristo, Dobson,
  and Vendruscolo]{Lindorff-Larsen2005}
Kresten Lindorff-Larsen, Robert~B Best, Mark~A DePristo, Christopher~M Dobson,
  and Michele. Vendruscolo.
\newblock {Simultaneous determination of protein structure and dynamics.}
\newblock \emph{Nature}, 433\penalty0 (7022):\penalty0 128--132, 2005.
\newblock ISSN 0028-0836.
\newblock \doi{10.1038/nature03199}.

\bibitem[White et~al.(2017)White, Knight, Hocky, and Voth]{White2017}
Andrew~D. White, Chris Knight, Glen~M. Hocky, and Gregory~A. Voth.
\newblock {Communication: Improved ab initio molecular dynamics by minimally
  biasing with experimental data}.
\newblock \emph{J. Chem. Phys.}, 146\penalty0 (4), 2017.
\newblock ISSN 00219606.
\newblock \doi{10.1063/1.4974837}.

\bibitem[Hummer and K{\"{o}}finger(2015)]{Hummer2015}
Gerhard Hummer and J{\"{u}}rgen K{\"{o}}finger.
\newblock {Bayesian ensemble refinement by replica simulations and
  reweighting}.
\newblock \emph{J. Chem. Phys.}, 143\penalty0 (24):\penalty0 243150, 2015.
\newblock ISSN 00219606.
\newblock \doi{10.1063/1.4937786}.

\bibitem[Marinelli and Faraldo-G{\'{o}}mez(2015)]{Marinelli2015}
Fabrizio Marinelli and Jos{\'{e}}~D. Faraldo-G{\'{o}}mez.
\newblock {Ensemble-Biased Metadynamics: A Molecular Simulation Method to
  Sample Experimental Distributions}.
\newblock \emph{Biophys. J.}, 108\penalty0 (12):\penalty0 2779--2782, 2015.
\newblock ISSN 15420086.
\newblock \doi{10.1016/j.bpj.2015.05.024}.

\bibitem[Gil-Ley et~al.(2016)Gil-Ley, Bottaro, and Bussi]{Gil-Ley2016}
Alejandro Gil-Ley, Sandro Bottaro, and Giovanni Bussi.
\newblock {Empirical Corrections to the Amber RNA Force Field with Target
  Metadynamics}.
\newblock \emph{J. Chem. Theory Comput.}, 12\penalty0 (6):\penalty0 2790--2798,
  2016.
\newblock ISSN 15499626.
\newblock \doi{10.1021/acs.jctc.6b00299}.

\bibitem[Valsson and Parrinello(2014)]{valsson2014variational}
Omar Valsson and Michele Parrinello.
\newblock Variational approach to enhanced sampling and free energy
  calculations.
\newblock \emph{Phys. Rev. Lett}, 113\penalty0 (9):\penalty0 090601, 2014.

\bibitem[Dama et~al.(2015)Dama, Hocky, Sun, and Voth]{Dama2015}
James~F. Dama, Glen~M. Hocky, Rui Sun, and Gregory~A. Voth.
\newblock {Exploring Valleys without Climbing Every Peak: More Efficient and
  Forgiving Metabasin Metadynamics via Robust On-the-Fly Bias Domain
  Restriction}.
\newblock \emph{J. Chem. Theory Comput.}, 11\penalty0 (12):\penalty0
  5638--5650, 2015.
\newblock ISSN 15499626.
\newblock \doi{10.1021/acs.jctc.5b00907}.

\bibitem[Pitera and Chodera(2012)]{Pitera2012}
Jed~W. Pitera and John~D. Chodera.
\newblock {On the Use of Experimental Observations to Bias Simulated
  Ensembles}.
\newblock \emph{J. Chem. Theory Comput.}, 8\penalty0 (10):\penalty0 3445--3451,
  2012.
\newblock ISSN 1549-9618.
\newblock \doi{10.1021/ct300112v}.

\bibitem[Jaynes(1957)]{PhysRev.106.620}
E~T Jaynes.
\newblock {Information Theory and Statistical Mechanics}.
\newblock \emph{Phys. Rev.}, 106\penalty0 (4):\penalty0 620--630, 1957.
\newblock \doi{10.1103/PhysRev.106.620}.

\bibitem[Roux and Weare(2013)]{Roux2013c}
Beno{\^{i}}t Roux and Jonathan Weare.
\newblock {On the statistical equivalence of restrained-ensemble simulations
  with the maximum entropy method}.
\newblock \emph{J. Chem. Phys.}, 138\penalty0 (8), 2013.
\newblock ISSN 00219606.
\newblock \doi{10.1063/1.4792208}.

\bibitem[Boura et~al.(2011)Boura, Rozycki, Herrick, Chung, Vecer, Eaton,
  Cafiso, Hummer, and Hurley]{Boura2011}
E.~Boura, B.~Rozycki, D.~Z. Herrick, H.~S. Chung, J.~Vecer, W.~A. Eaton, D.~S.
  Cafiso, G.~Hummer, and J.~H. Hurley.
\newblock {Solution structure of the ESCRT-I complex by small-angle X-ray
  scattering, EPR, and FRET spectroscopy}.
\newblock \emph{Proc. Natl. Acad. Sci.}, 108\penalty0 (23):\penalty0
  9437--9442, 2011.
\newblock ISSN 0027-8424.
\newblock \doi{10.1073/pnas.1101763108}.

\bibitem[McMahan and Streeter(2010)]{McMahan2010}
HB~McMahan and Matthew Streeter.
\newblock {Adaptive bound optimization for online convex optimization}.
\newblock \emph{arXiv}, pages 1--19, 2010.
\newblock ISSN 09505849.
\newblock \doi{10.1016/j.infsof.2008.09.005}.

\bibitem[Hocky et~al.(2017)Hocky, Dannenhoffer-Lafage, and Voth]{Hocky2017}
Glen~M. Hocky, Thomas Dannenhoffer-Lafage, and Gregory~A. Voth.
\newblock {Coarse-Grained Directed Simulation}.
\newblock \emph{J. Chem. Theory Comput.}, 13\penalty0 (9):\penalty0 4593--4603,
  2017.
\newblock ISSN 15499626.
\newblock \doi{10.1021/acs.jctc.7b00690}.

\bibitem[Dama et~al.(2014)Dama, Parrinello, and Voth]{Dama2014a}
James~F. Dama, Michele Parrinello, and Gregory~a. Voth.
\newblock {Well-Tempered Metadynamics Converges Asymptotically}.
\newblock \emph{Phys. Rev. Lett.}, 112\penalty0 (24):\penalty0 240602, 2014.
\newblock ISSN 0031-9007.
\newblock \doi{10.1103/PhysRevLett.112.240602}.

\bibitem[Brookes and Head-Gordon(2016)]{Brookes2016}
David~H. Brookes and Teresa Head-Gordon.
\newblock {Experimental Inferential Structure Determination of Ensembles for
  Intrinsically Disordered Proteins}.
\newblock \emph{J. Am. Chem. Soc.}, 138\penalty0 (13):\penalty0 4530--4538,
  2016.
\newblock ISSN 15205126.
\newblock \doi{10.1021/jacs.6b00351}.

\bibitem[Cesari et~al.(2016)Cesari, Gil-Ley, and Bussi]{Cesari2016}
Andrea Cesari, Alejandro Gil-Ley, and Giovanni Bussi.
\newblock {Combining Simulations and Solution Experiments as a Paradigm for RNA
  Force Field Refinement}.
\newblock \emph{J. Chem. Theory Comput.}, 12\penalty0 (12):\penalty0
  6192--6200, 2016.
\newblock ISSN 15499626.
\newblock \doi{10.1021/acs.jctc.6b00944}.

\bibitem[Skinner et~al.(2013)Skinner, Huang, Schlesinger, Pettersson, Nilsson,
  and Benmore]{Skinner2013}
Lawrie~B. Skinner, Congcong Huang, Daniel Schlesinger, Lars~G.M. Pettersson,
  Anders Nilsson, and Chris~J. Benmore.
\newblock {Benchmark oxygen-oxygen pair-distribution function of ambient water
  from x-ray diffraction measurements with a wide Q-range}.
\newblock \emph{J. Chem. Phys.}, 138\penalty0 (7), 2013.
\newblock ISSN 00219606.
\newblock \doi{10.1063/1.4790861}.

\bibitem[Tse et~al.(2015)Tse, Knight, and Voth]{Tse2015}
Ying Lung~Steve Tse, Chris Knight, and Gregory~A. Voth.
\newblock {An analysis of hydrated proton diffusion in ab initio molecular
  dynamics}.
\newblock \emph{J. Chem. Phys.}, 142\penalty0 (1), 2015.
\newblock ISSN 00219606.
\newblock \doi{10.1063/1.4905077}.

\bibitem[Krynicki et~al.(1978)Krynicki, Green, and Sawyer]{DC9786600199}
Kazimierz Krynicki, Christopher~D Green, and David~W Sawyer.
\newblock {Pressure and temperature dependence of self-diffusion in water}.
\newblock \emph{Faraday Discuss. Chem. Soc.}, 66\penalty0 (0):\penalty0
  199--208, 1978.
\newblock \doi{10.1039/DC9786600199}.

\bibitem[Sitarz et~al.(2000)Sitarz, Wirth-Dzieciolowska, and
  Demant]{Sitarz2000}
M.~Sitarz, E.~Wirth-Dzieciolowska, and P.~Demant.
\newblock {Loss of heterozygosity on chromosome 5 in vicinity of the telomere
  in $\gamma$-radiation-induced thymic lymphomas in mice}.
\newblock \emph{Neoplasma}, 47\penalty0 (3):\penalty0 148--150, 2000.
\newblock ISSN 00282685.
\newblock \doi{10.1063/1.3382344}.

\bibitem[Cortina et~al.(2018)Cortina, Hays, and Kasson]{Cortina2018}
George~A. Cortina, Jennifer~M. Hays, and Peter~M. Kasson.
\newblock {Conformational Intermediate That Controls KPC-2 Catalysis and
  Beta-Lactam Drug Resistance}.
\newblock \emph{ACS Catal.}, 8\penalty0 (4):\penalty0 2741--2747, 2018.
\newblock ISSN 21555435.
\newblock \doi{10.1021/acscatal.7b03832}.

\bibitem[Pemberton et~al.(2017)Pemberton, Zhang, and
  Chen]{doi:10.1021/acs.jmedchem.7b00158}
Orville~A Pemberton, Xiujun Zhang, and Yu~Chen.
\newblock {Molecular Basis of Substrate Recognition and Product Release by the
  Klebsiella pneumoniae Carbapenemase (KPC-2)}.
\newblock \emph{J. Med. Chem}, 60\penalty0 (8):\penalty0 3525--3530, 2017.
\newblock \doi{10.1021/acs.jmedchem.7b00158}.

\bibitem[Du et~al.(1998)Du, Pande, Grosberg, Tanaka, and
  Shakhnovich]{du1998transition}
Rose Du, Vijay~S Pande, Alexander~Yu Grosberg, Toyoichi Tanaka, and Eugene~S
  Shakhnovich.
\newblock {On the transition coordinate for protein folding}.
\newblock \emph{J. Chem. Phys.}, 108\penalty0 (1):\penalty0 334--350, 1998.

\bibitem[Dannenhoffer-Lafage et~al.(2016)Dannenhoffer-Lafage, White, and
  Voth]{Dannenhoffer-Lafage2016}
Thomas Dannenhoffer-Lafage, Andrew~D White, and Gregory~A Voth.
\newblock {A Direct Method for Incorporating Experimental Data into Multiscale
  Coarse-grained Models.}
\newblock \emph{J. Chem. Theory Comput.}, 12\penalty0 (5):\penalty0 2144--2153,
  2016.
\newblock ISSN 1549-9626.
\newblock \doi{10.1021/acs.jctc.6b00043}.

\bibitem[Masia et~al.(2004)Masia, Probst, and Rey]{Masia2004}
Marco Masia, Michael Probst, and Rossend. Rey.
\newblock {Ethylene Carbonate-Li+: A Theoretical Study of Structural and
  Vibrational Properties in Gas and Liquid Phases.}
\newblock \emph{J. Phys. Chem. B}, 108\penalty0 (6):\penalty0 2016--2027, 2004.
\newblock ISSN 1520-6106.
\newblock \doi{10.1021/jp036673w}.

\bibitem[Borodin and Smith(2009)]{Borodin2009}
Oleg Borodin and Grant~D Smith.
\newblock {Quantum Chemistry and Molecular Dynamics Simulation Study of
  Dimethyl Carbonate: Ethylene Carbonate Electrolytes Doped with LiPF6.}
\newblock \emph{J. Phys. Chem. B}, 113\penalty0 (6):\penalty0 1763--1776, 2009.
\newblock ISSN 1520-6106.
\newblock \doi{10.1021/jp809614h}.

\bibitem[Wagner et~al.(2016)Wagner, Dama, Durumeric, and Voth]{Wagner2016b}
Jacob~W. Wagner, James~F. Dama, Aleksander~E.P. Durumeric, and Gregory~A. Voth.
\newblock {On the representability problem and the physical meaning of
  coarse-grained models}.
\newblock \emph{J. Chem. Phys.}, 145\penalty0 (4), 2016.
\newblock ISSN 00219606.
\newblock \doi{10.1063/1.4959168}.

\bibitem[Stinis(2005)]{Stinis2005}
Panagiotis Stinis.
\newblock {A maximum likelihood algorithm for the estimation and
  renormalization of exponential densities}.
\newblock \emph{J. Comput. Phys.}, 208\penalty0 (2):\penalty0 691--703, 2005.
\newblock ISSN 00219991.
\newblock \doi{10.1016/j.jcp.2005.03.001}.

\bibitem[Tiwary and Parrinello(2015)]{Tiwary2015}
Pratyush Tiwary and Michele Parrinello.
\newblock {A Time-Independent Free Energy Estimator for Metadynamics}.
\newblock \emph{J. Phys. Chem. B}, 119\penalty0 (3):\penalty0 736--742, 2015.
\newblock ISSN 1520-6106.
\newblock \doi{10.1021/jp504920s}.

\bibitem[Amirkulova and White(2018)]{Amirkulova2018}
Dilnoza~B. Amirkulova and Andrew~D. White.
\newblock {Combining enhanced sampling with experiment-directed simulation of
  the GYG peptide}.
\newblock \emph{J. Theor. Comput. Chem.}, 17\penalty0 (03):\penalty0 1840007,
  2018.
\newblock ISSN 0219-6336.
\newblock \doi{10.1142/S0219633618400072}.

\bibitem[Bonomi and Parrinello(2010)]{Bonomi2010}
M.~Bonomi and M.~Parrinello.
\newblock {Enhanced sampling in the well-tempered ensemble}.
\newblock \emph{Phys. Rev. Lett.}, 104\penalty0 (19):\penalty0 1--4, 2010.
\newblock ISSN 00319007.
\newblock \doi{10.1103/PhysRevLett.104.190601}.

\bibitem[Deighan et~al.(2012)Deighan, Bonomi, and Pfaendtner]{Deighan2012a}
Michael Deighan, Massimiliano Bonomi, and Jim Pfaendtner.
\newblock {Efficient simulation of explicitly solvated proteins in the
  well-tempered ensemble}.
\newblock \emph{J. Chem. Theory Comput.}, 8\penalty0 (7):\penalty0 2189--2192,
  2012.
\newblock ISSN 15499618.
\newblock \doi{10.1021/ct300297t}.

\bibitem[Ting et~al.(2010)Ting, Wang, Shapovalov, Mitra, Jordan, and
  Dunbrack]{Ting2010}
Daniel Ting, Guoli Wang, Maxim Shapovalov, Rajib Mitra, Michael~I. Jordan, and
  Roland~L. Dunbrack.
\newblock {Neighbor-dependent Ramachandran probability distributions of amino
  acids developed from a hierarchical dirichlet process model}.
\newblock \emph{PLOS Comput. Biol.}, 6\penalty0 (4), 2010.
\newblock ISSN 1553734X.
\newblock \doi{10.1371/journal.pcbi.1000763}.

\bibitem[Louwerse and Baerends(2006)]{Louwerse2006}
Manuel~J. Louwerse and Evert~Jan Baerends.
\newblock {Calculation of pressure in case of periodic boundary conditions}.
\newblock \emph{Chem. Phys. Lett.}, 421\penalty0 (1-3):\penalty0 138--141,
  2006.
\newblock ISSN 00092614.
\newblock \doi{10.1016/j.cplett.2006.01.087}.

\bibitem[Plimpton(1995)]{Plimpton1995}
Steve Plimpton.
\newblock Fast parallel algorithms for short-range molecular dynamics.
\newblock \emph{J. Comput. Phys.}, 117\penalty0 (1):\penalty0 1 -- 19, 1995.
\newblock ISSN 0021-9991.
\newblock \doi{https://doi.org/10.1006/jcph.1995.1039}.

\bibitem[Scholkopf and Smola(2001)]{scholkopf2001learning}
Bernhard Scholkopf and Alexander~J Smola.
\newblock \emph{Learning with kernels: support vector machines, regularization,
  optimization, and beyond}.
\newblock MIT press, 2001.

\bibitem[Fiorin et~al.(2013)Fiorin, Klein, and H{\'{e}}nin]{COLVARS}
Giacomo Fiorin, Michael~L Klein, and J{\'{e}}r{\^{o}}me H{\'{e}}nin.
\newblock {Using collective variables to drive molecular dynamics simulations}.
\newblock \emph{Mol. Phys.}, 111\penalty0 (22-23):\penalty0 3345--3362, 2013.
\newblock \doi{10.1080/00268976.2013.813594}.

\bibitem[Bonomi et~al.(2009)Bonomi, Branduardi, Bussi, Camilloni, Provasi,
  Raiteri, Donadio, Marinelli, Pietrucci, Broglia, and Parrinello]{PLUMED}
Massimiliano Bonomi, Davide Branduardi, Giovanni Bussi, Carlo Camilloni, Davide
  Provasi, Paolo Raiteri, Davide Donadio, Fabrizio Marinelli, Fabio Pietrucci,
  Ricardo~A Broglia, and Michele Parrinello.
\newblock {{\{}PLUMED{\}}: A portable plugin for free-energy calculations with
  molecular dynamics}.
\newblock \emph{Comput. Phys. Commun.}, 180\penalty0 (10):\penalty0 1961--1972,
  2009.
\newblock ISSN 00104655.
\newblock \doi{10.1016/j.cpc.2009.05.011}.

\bibitem[Tribello et~al.(2014)Tribello, Bonomi, Branduardi, Camilloni, and
  Bussi]{TRIBELLO2014604}
Gareth~A Tribello, Massimiliano Bonomi, Davide Branduardi, Carlo Camilloni, and
  Giovanni Bussi.
\newblock {PLUMED 2: New feathers for an old bird}.
\newblock \emph{Comput. Phys. Commun}, 185\penalty0 (2):\penalty0 604--613,
  2014.
\newblock ISSN 0010-4655.
\newblock \doi{https://doi.org/10.1016/j.cpc.2013.09.018}.

\bibitem[Bussi and Tribello(2018)]{bussi2018analyzing}
Giovanni Bussi and Gareth~A Tribello.
\newblock Analyzing and biasing simulations with plumed.
\newblock \emph{arXiv}, 2018.

\end{thebibliography}

\appendix

\section{Minimal Biasing Pressure Derivation}

We would like to constrain the virial contribution to pressure while minimizing relative entropy in a biased ensemble. The constraints are:

\begin{equation}
\int d\vec{r}\, P'(\vec{r}) = 1
\end{equation}

which enforces normalization. The other constraint is to set the virial pressure, $V[P'(\vec{r})]$, to be equal to the scalar $p_e$.

\begin{equation}
\label{eq:virial}
V[P'(\vec{r})] = \frac{1}{3V}\int d\vec{r}\, \vec{r}\cdot  \frac{\partial \ln P'(\vec{r})}{\partial \vec{r}} P'(\vec{r}) =  \int d\vec{r}\, \vec{r}\cdot  \nabla P'(\vec{r}) = p_e
\end{equation}

where $V$ is the volume. The quantity to be minimized is relative entropy of the biased probability distribution with respect to the unbiased probability distribution:

\begin{equation}
\int P(\vec{r}) \ln\frac{P'(\vec{r})}{P(\vec{r})}
\end{equation}

To solve this system, we compute the functional derivative of the Lagrangian with respect to the biased probability distribution $P'(\vec{r})$. This requires a functional derivative of Equation~\ref{eq:virial}, which is:

\begin{equation}
\label{eq:virial-deriv}
\frac{\Delta p[P'(\vec{r})]}{\delta P'(\vec{r})} = - \nabla \left[r_1, r_2,...\right] = -N
\end{equation}

Some intuition for Equation~\ref{eq:virial-deriv} can be gained from the definition of the functional derivative. Consider adding an arbitrary function $\phi(\vec{r})$ to the probability distribution scaled by $\epsilon$, where $\epsilon$ is small. The change in the virial with this additional function is

\begin{equation}
\int d\vec{r}\, \vec{r}\cdot  \nabla \left[P(\vec{r}) + \epsilon \phi(\vec{r})\right] -  \int d\vec{r}\, \vec{r}\cdot  \nabla P(\vec{r}) \approx -N \int dr\, \epsilon\phi(\vec{r})
\end{equation}

The right-hand side is independent of the probability distribution, so that changing the probability distribution decreases the virial by $N$ regardless of where the change occurs. With the functional derivative of the virial being constant, the functional derivative of the Lagrangian is

\begin{equation}
\label{eq:virial-lagrange}
\frac{\delta \mathcal{L}}{\delta P'(\vec{r})} = \ln P'(\vec{r}) - \ln P(\vec{r}) + \lambda_1 - \lambda_2 N
\end{equation}

which has no solution for $P'(\vec{r})$ because $\lambda_1$ and $\lambda_2$ are colinear. Thus there is no way to constrain pressure. 

\end{document}